\documentclass[conference]{IEEEtran}
\IEEEoverridecommandlockouts

\usepackage{cite}
\usepackage{amsmath,amssymb,amsfonts}
\usepackage{algorithmic}
\usepackage{graphicx}
\usepackage{textcomp}
\usepackage{enumitem}
\usepackage{url}
\usepackage{xcolor}
\usepackage{orcidlink}
\usepackage{booktabs}
\usepackage{hyperref}
\def\BibTeX{{\rm B\kern-.05em{\sc i\kern-.025em b}\kern-.08em
    T\kern-.1667em\lower.7ex\hbox{E}\kern-.125emX}}
\begin{document}

\title{Virtual Reality-Based Telerehabilitation for
Upper Limb Recovery Post-Stroke: A Systematic
Review of Design Principles, Monitoring, Safety,
and Engagement Strategies}

\author{Pedro Rodrigues~\orcidlink{0009-0002-0232-3017}, Cláudia Quaresma~\orcidlink{0000-0001-9978-261X}, Maria Costa, Filipe Luz~\orcidlink{0000-0002-3608-8417}, and Maria Fonseca~\orcidlink{0000-0001-7946-4825}
\thanks{This work is supported by Fundação para a Ciência e Tecnologia (FCT), under LIBPhys \url{https://doi.org/10.54499/UIDB/04559/2020}, HEI-Lab R\&D Unit \url{https://doi.org/10.54499/UIDB/05380/2020}, and Project GameIn: Games Inclusion Lab: Participatory Media Creation Processes for Accessibility \url{https://doi.org/10.54499/2022.07939.PTDC}. (Corresponding author: Cláudia Quaresma) }
\thanks{Pedro Rodrigues is with the Laboratory for Instrumentation, Biomechanical Engineering and Radiation Physics, Faculty of Sciences and Technology, NOVA University Lisbon, 2829-516 Caparica, Portugal (e-mail: pme.rodrigues@campus.fct.unl.pt)}
\thanks{Cláudia Quaresma is with the Laboratory for Instrumentation, Biomechanical Engineering and Radiation Physics, Faculty of Sciences and Technology, NOVA University Lisbon, 2829-516 Caparica, Portugal, and also the Physics Department, Faculty of Sciences and Technology, NOVA University Lisbon, 2829-516 Caparica, Portugal (e-mail: q.claudia@fct.unl.pt)}
\thanks{Maria Costa and Filipe Luz are with HEI-Lab Digital Human-environment Interaction Lab, Lusófona University, Avenida do Campo Grande, 1749-024 Lisboa (e-mail: mariaqgcosta2001@gmail.com; filipe.luz@ulusofona.pt)}
\thanks{Maria Fonseca is with the Laboratory for Instrumentation, Biomechanical Engineering and Radiation Physics, Faculty of Sciences and Technology, NOVA University Lisbon, 2829-516 Caparica, Portugal, and also with HEI-Lab Digital Human-environment Interaction Lab, Lusófona University, Avenida do Campo Grande, 1749-024 Lisboa (e-mail: micaela.fonseca@ulusofona.pt).}
}

\maketitle

\begin{abstract}
Stroke rehabilitation continues to face challenges in accessibility and patient engagement, where traditional approaches often fall short. Virtual reality (VR)-based telerehabilitation offers a promising avenue, by enabling home-based recovery through immersive environments and gamification. This systematic review evaluates current VR solutions for upper-limb post-stroke recovery, focusing on design principles, safety measures, patient-therapist communication, and strategies to promote motivation and adherence. Following PRISMA 2020 guidelines, a comprehensive search was conducted across PubMed, IEEE Xplore, and ScienceDirect. The review reveals a scarcity of studies meeting the inclusion criteria, possibly reflecting the challenges inherent in the current paradigm of VR telerehabilitation systems. Although these systems have potential to enhance accessibility and patient autonomy, they often lack standardized safety protocols and reliable real-time monitoring. Human-centered design principles are evident in some solutions, but inconsistent patient involvement during the development process limits their usability and clinical relevance. Furthermore, communication between patients and therapists remains constrained by technological barriers, although advancements in real-time feedback and adaptive systems offer promising solutions. This review underscores the potential of VR telerehabilitation to address critical needs in upper-limb stroke recovery while highlighting the importance of addressing existing limitations to ensure broader clinical implementation and improved patient outcomes.

\end{abstract}

\begin{IEEEkeywords}
Human-centered design, Patient engagement, Safety standards, Stroke, Telerehabilitation, Upper limb, Virtual reality
\end{IEEEkeywords}

\section{Introduction}
\label{sec:introduction}

\IEEEPARstart{S}{troke} is a major global health concern, ranking as the second leading cause of death worldwide and expected to affect one in four individuals over 25 during their lifetime \cite{WSO}. Between 1990 and 2019, stroke-related deaths increased by 43\%, accounting for approximately 11\% of global deaths in 2019 \cite{WSO}. It is also the third leading cause of combined death and disability, with disability-adjusted life years (DALYs) significantly rising during this period \cite{WSO}. This rise underscores a growing burden as more individuals either die prematurely or live with long-term disabilities. Upper limb motor impairments are prevalent among stroke survivors, impacting around 90\% of acute patients and persisting in up to 60\% of chronic cases \cite{RefGuide, Kwakkel}. These impairments lead to increased dependency and contribute to declines in physical, psychological, social, and emotional well-being \cite{Lazem2023}. Thus, upper limb recovery is essential for promoting patient independence, self-efficacy, and quality of life \cite{Lazem2023}. However, traditional rehabilitation services often fall short, especially for individuals in remote areas \cite{kim2022}.

Telerehabilitation leverages information and communication technologies to provide home-based care with remote therapist monitoring, bridging the gap between patients and healthcare providers \cite{davidbrennan}. It offers a cost-effective alternative to clinic visits, particularly benefiting remote patients through flexible, patient-centered scheduling \cite{janefonseca}. Studies indicate that telerehabilitation can be as effective as face-to-face rehabilitation in improving physical function, while adhering to equal principles like goal setting, education, and social support \cite{Leslie2021, Lazem2023}.

Advancements in health technologies have expanded telerehabilitation to include mobile applications, games, sensors, and robotics \cite{sensors}. Among these innovations, virtual reality (VR) stands out, offering rehabilitation through immersive simulated environments \cite{jie2023}. Utilizing head-mounted displays and controllers, VR provides real-time feedback and multi-sensory interaction, promoting motivation and enjoyment while enabling task customization for more effective treatment \cite{Tieri2017, fregna2023}. VR also facilitates detailed data tracking \cite{dejaco2023} and promotes motor learning and neural plasticity, leading to better skill retention and real-world transfer \cite{jie2022}. Combined with the benefits of telerehabilitation, VR-based approaches have the potential to improve healthcare access, data analysis, patient motivation, and adherence.

However, the efficacy of VR in rehabilitating upper limb motor impairments remains debatable. A recent systematic review \cite{silvia2023} found that VR rehabilitation generally outperforms conventional therapy in improving upper limb function post-stroke, as measured by the Fugl-Meyer Assessment, but evidence for other measures is inconsistent. These mixed outcomes indicate that while VR holds potential, clinicians should remain aware of the current limitations in the evidence.

This underscores the need for robust design principles and best practices in human-centered VR research \cite{brandon2019}. The lack of comprehensive guidelines in VR health studies highlights the importance of evaluating current solutions holistically, beyond efficacy, to inform future developments \cite{brandon2019, lorenz2024}.

Despite advancements, the use of VR-based upper-limb rehabilitation in home settings remains underexplored, and systematic reviews addressing their multidimensional aspects are scarce. We posit that understanding key characteristics beyond effectiveness is critical to identify factors driving success. This review addresses these gaps by exploring four critical research dimensions through the following questions:

\vspace{0.15cm}

\begin{itemize}[label=\stepcounter{enumi}\textbf{Q\_}{\theenumi}]
    \item \textit{What are the key characteristics and design principles of VR-based telerehabilitation systems? }
    \item \textit{How is monitoring and communication facilitated during treatment? }
    \item \textit{How is safety addressed in VR telerehabilitation solutions? }
    \item \textit{What strategies are employed to enhance patient motivation, engagement, and satisfaction with VR telerehabilitation technologies? }
\end{itemize}

\section{Background}
\label{sec:background}

This section provides an overview of the four research dimensions employed in our review, and which we deem central to understanding the challenges and opportunities in telerehabilitation and VR-based rehabilitation. Section \ref{subsec:HCD} explores Human-Centered Design (HCD) practices, particularly within VR clinical studies. Section \ref{subsec:Accessibility and Communication} examines the accessibility and communication factors that condition therapists’ and patients’ adoption of telerehabilitation. Section \ref{subsec:Safety} overviews critical safety risks and precautions inherent to remote and VR environments. Finally, Section \ref{subsec:Motivation and Adherence} discusses how VR and game-based approaches can promote patient motivation and adherence, and support long-term rehabilitation goals.

\subsection{Human Centered Design}
\label{subsec:HCD}

HCD is a problem-solving approach that prioritizes end-users' needs, preferences, and limitations throughout the design lifecycle \cite{irene2021}. By fostering empathy and deep user understanding, HCD aims to create functional, user-friendly, and accessible designs, promoting user control, empowerment, and commitment, which increases the likelihood of adoption and sustained use \cite{irene2021, lorenz2024, maier2019}.

The application of HCD in scientific research, particularly in health-related fields, is growing, with promising examples in depression management, motor coordination therapy, clinical decision support systems, exergames, and autism interventions \cite{irene2021}.

VR technology presents opportunities to refine and expand HCD methods. Birckhead et al. \cite{brandon2019} propose an HCD framework tailored for VR health interventions, addressing the unique challenges of VR-based treatments. This approach leverages VR’s immersive and interactive nature to improve the design process, aligning systems more closely with user needs and optimizing usability and effectiveness \cite{brandon2019}.

They systematized their approach into guidelines across three development phases (VR1-3). The VR1 phase focuses on content creation through close collaboration with patients and healthcare providers, highlighting strategies like "inspiration through empathy" (understanding patients' experiences), "ideation through team collaboration" (brainstorming and prototyping), and "prototyping through continuous user feedback" (iterative testing and refinement).

In this context, HCD becomes critical for identifying and addressing adoption barriers faced by stakeholders, such as digital literacy challenges, lack of personal interaction, and safety concerns \cite{brandon2019, tyagi2017}. Although VR technology has significant potential to promote patient engagement, its novelty and complexity can lead to feelings of alienation, hindering adoption \cite{alessia2024}. Therefore, incorporating HCD can help make VR technology more accessible, user-friendly, and appealing for end-users \cite{irene2021}.

\subsection{Accessibility and Communication}
\label{subsec:Accessibility and Communication}

In addition to patient-related barriers, the adoption of telerehabilitation interventions by therapists involves a variety of facilitators and obstacles. The flexibility and accessibility of telerehabilitation allow for continuous patient care and enable therapists to reach a broader patient base. However, therapists are often more cautious than patients about incorporating technology into their established rehabilitation routines \cite{lorenz2024}.

The effectiveness of telerehabilitation relies heavily on therapists' ability to monitor, instruct, and correct patients during treatments. Without robust methods for qualitative and quantitative performance measurement, there is a risk of overtraining and potential injuries, as patients may overestimate their capabilities \cite{cho2022}. Concerns about the effectiveness of telerehabilitation, especially compared to face-to-face interventions for complex cases, still persist among healthcare providers \cite{niknejad2021}.

Therapists also face significant challenges in adapting to new technologies, including a steep learning curve and limited time to familiarize themselves with these systems \cite{niknejad2021}. The time required to learn and set up new technologies can hinder their adoption \cite{lorenz2024}. Additionally, concerns about managing technology-related issues \cite{niknejad2021}, such as unreliable internet connections, insufficient technical support, data security, and patient confidentiality further delay the integration of telerehabilitation into routine practice \cite{hale-gallargo2020}.

Thus, transitioning rehabilitation interventions into settings, such as homes, presents additional challenges, particularly in providing adequate access to technology and support systems. The lack of strong communication channels and support structures can hinder progress monitoring and the effective exchange of information with key stakeholders, creating further barriers to the successful integration and adoption of these interventions in clinical practice.

\subsection{Safety}
\label{subsec:Safety}

In any form of treatment, patient safety and well-being are paramount. While telerehabilitation and VR rehabilitation offer numerous benefits, they also carry risks, including an increased potential for injury. In telerehabilitation, the absence of direct physical supervision can lead to musculoskeletal injuries, as patients may perform exercises incorrectly or overexert themselves \cite{buckingham2022}. Although the ability to exercise independently is crucial for the transition from inpatient to outpatient care, unmonitored exercises pose a higher risk, especially in the early stages of rehabilitation when poor workload management and overconfidence are common issues \cite{lorenz2024}.

Additionally, virtual assessments also have limitations that can result in inadequate evaluations of a patient’s physical capabilities, potentially leading to inappropriate exercise prescriptions \cite{jirasakulsuk2022}. The lack of immediate physical intervention during remote sessions means therapists cannot promptly address discomfort or adverse reactions, posing another significant risk to patient safety \cite{cox2021}. Technological issues, such as internet connectivity problems, can disrupt sessions, causing frustration and potentially leading patients to perform exercises without proper guidance \cite{niknejad2021}.

In VR rehabilitation, similar risks are present. The immersive and often intense nature of VR exercises can lead to visual and physical strain, as well as injuries. Users may experience motion sickness or dizziness, which can deter consistent use and increase the risk of falls or accidents \cite{bui2021}. Moreover, home-based VR training often takes place in unsupervised and suboptimal environments, which can further heighten the risk of injury \cite{lorenz2024}.

\subsection{Motivation and Adherence}
\label{subsec:Motivation and Adherence}

Traditional rehabilitation often faces challenges with patient adherence, particularly in home settings where patients exercise alone. The lack of support, real-time feedback, and therapist presence can significantly undermine patients' willingness to engage in exercises independently \cite{rob2018}. This absence of motivation can lead to passivity and stagnation in progress, further hindering recovery \cite{lorenz2024}.

Motivation is a critical component in rehabilitation and a primary focus for therapists \cite{lorenz2024}. It is a psychological trait that drives individuals to initiate and persist in goal-oriented tasks, often without the need for constant encouragement. Intrinsic motivation, which comes from genuine interest and enjoyment in an activity, is particularly valuable as it fosters a sense of fulfillment and sustained engagement in therapy \cite{lorenz2024}. Therefore, rehabilitation systems must not only be tailored to user needs but also designed to be intrinsically motivating, thereby reducing reliance on therapists for motivation.

Game-based therapies have become popular in rehabilitation due to their ability to enhance intrinsic motivation. These games are designed to align with patients' therapeutic needs while introducing short-term goals and offering immediate rewards. This immediate feedback and sense of accomplishment make the rehabilitation process more enjoyable and engaging \cite{keith2013}. However, many of these games involve simple tasks on a computer screen, which can feel disconnected from everyday gestures and movements \cite{fregna2022}. This disconnect may limit their effectiveness in transferring motor skills to real-life scenarios \cite{maier2019}.

In this context, VR offers a significant advantage for context-based training. For example, VR can simulate settings where patients practice interactions, such as driving a car or using household objects, allowing them to rehearse movements in a semi-realistic context. The realistic proprioceptive feedback provided by VR helps bridge the gap between therapy and everyday life, facilitating the transfer of skills to real-world activities \cite{fregna2023}.

Nevertheless, technology alone is not sufficient to sustain continuous motivation, as motivation can fluctuate due to various factors, such as the diminishing novelty of an activity or slow progress. The lack of immediate professional feedback and support in home settings can further impact motivation levels.

While the literature offers various methods for assessing motivation, such as surveys and interviews \cite{lorenz2024}, effective rehabilitation must also consider individual factors like age, pathology, personal preferences, and goals \cite{lorenz2024}. Personal relationships and interactions are also significant motivational drivers, along with patients' interests and hobbies \cite{lorenz2024}. Consequently, treatment journeys should account for these factors, by incorporating individual perspectives and concerns, setting initial goals, and adapting these goals throughout the rehabilitation process.

\section{Methodology}
\label{sec:methodology}

\subsection{Study Design}

The present systematic review was conducted following the Preferred Reporting Items for Systematic Reviews and Meta-Analyses (PRISMA) guidelines \cite{prisma2020}. The PRISMA checklist was used to ensure comprehensive and transparent reporting throughout the review process.

\subsection{Eligibility Criteria}

The review used a mixed-methods approach, incorporating both quantitative and qualitative research. Table~\ref{tab:criteria} provides an overview of the various criteria. Studies not meeting these criteria were excluded from the final review.

\begin{table}[]
\footnotesize
\caption{Overview of Inclusion and Exclusion Criteria}
\label{tab:criteria}
\begin{tabular}{p{1.6cm} p{3cm} p{3cm}}
\toprule
\textbf{Criteria} & \textbf{Included} & \textbf{Excluded} \\
\midrule
\textbf{Methodologies Included} & Mixed-methods: RCTs, non-RCTs, observational studies (cohort, case-control, cross-sectional), qualitative studies, mixed-methods studies, case studies, pilot, and feasibility studies & Reviews, meta-analyses, protocols, books, studies without primary data or clear methodology, editorials \\
\midrule
\textbf{Participants} & Adults aged 18+ with upper motor impairments, specifically post-stroke patients & Populations without upper limb motor disorders, conditions related to Alzheimer’s, Parkinson’s, epilepsy, multiple sclerosis \\
\midrule
\textbf{Interventions} & VR technologies for telerehabilitation, designed for upper motor function rehabilitation & Non-VR interventions, VR applications not designed for telerehabilitation or upper motor function rehabilitation \\
\midrule
\textbf{Outcomes} & Effectiveness, usability, patient engagement, quality of life, safety, satisfaction & Outcomes unrelated to upper motor function rehabilitation, solely technical aspects of VR development \\
\midrule
\textbf{Language} & Studies written in English & Studies published in other languages without available translations \\
\midrule
\textbf{Publication Type \& Date} & Peer-reviewed articles published from 2005 onwards & Abstracts, conference proceedings without full text, patents, books \\
\bottomrule
\end{tabular}
\end{table}

\subsection{Search Strategy}
A comprehensive literature search was conducted in February 2024, using the electronic databases PubMed, Scopus, IEEE Xplore, ACM, and Web of Science. The databases were selected based on their relevance to our research goals, as informed by prior reviews that explored similar topics \cite{Lazem2023, jie2023, moulaei2022, gebreheat2024}. The search focused on studies published from 2005 onwards. The keywords for the systematic review were selected and tested iteratively via preliminary searches across the various databases to ensure they aligned with the inclusion criteria. This method refined the keyword string, by balancing the volume of the search results with their relevance. The keywords used in the final search string were organized into four categories: (1) "telerehabilitation", (2) "virtual reality", (3) "post-stroke", (4) "upper limb". Each term included additional related terms combined using the appropriate boolean (AND/OR) and wildcard (*) operators to maximize the retrieval of relevant results.

\subsection{Reviewing Procedures}

Two reviewers independently conducted the initial screening of sources based on titles and abstracts. The review process was managed using Rayyan software \cite{rayyan}. Eligible studies were then downloaded in full text and assessed for inclusion according to predefined criteria. Any conflicts during this assessment were resolved through discussion with a third reviewer. An overview of the reviewing process is depicted in the PRISMA flow chart (figure \ref{fig:flowdiagram}).

\section{Results}

\begin{figure}[t]
    \centering
    \includegraphics[width=\columnwidth, trim=0 60 0 0, clip]{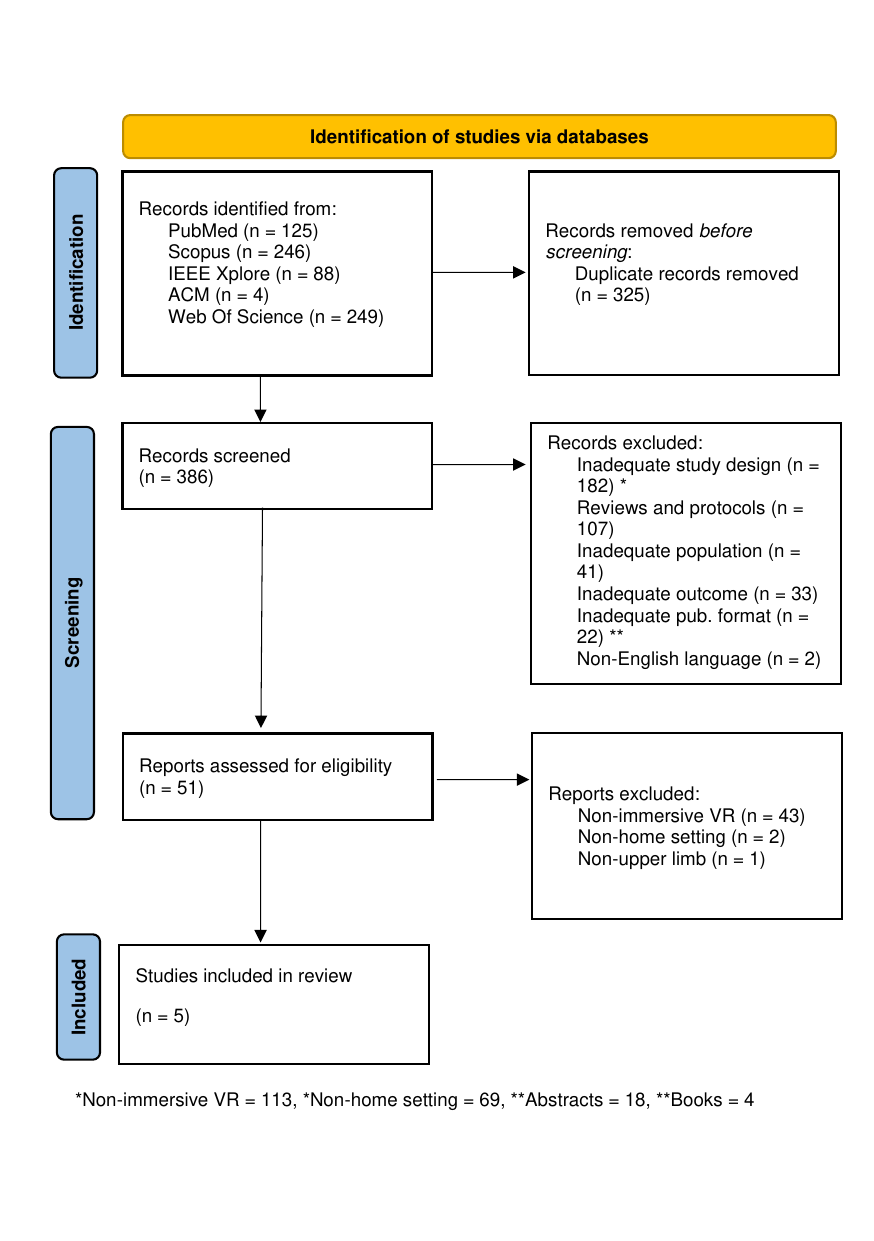}
    \caption{PRISMA 2020 flow diagram.}
    \label{fig:flowdiagram}
\end{figure}

The search process initially yielded 712 results, from which 325 duplicates were removed. After screening titles and abstracts, 336 studies were excluded, leaving 51 articles for full-text review. Ultimately, only 5 studies were selected for inclusion in the final review. A significant factor in the exclusion of many studies was the heterogeneity in defining "\textit{virtual reality}". Often, "VR" was used interchangeably with video games, which complicated our analysis. VR was specifically defined in this review by its use of head-mounted displays and specific features like immersivity and proprioception. As a result, only technologies meeting these criteria were considered VR and included, leading to the exclusion of 113 studies. Other major exclusion reasons included the involvement of non-target populations, such as healthy individuals or non-stroke patients, and technologies that did not focus on telerehabilitation or home solutions, resulting in the exclusion of 110 articles. A detailed overview of the exclusion reasons is provided in figure \ref{fig:flowdiagram}.

Key characteristics extracted from the selected studies include authorship, publication year, country, study type, objectives, and population details such as sample size, age, and time post-stroke. Intervention details were also collected, covering the setting, delivery mode, technologies used, and outcomes. In line with the review’s focus, data on stakeholder roles, their involvement in the research, and their influence on solution design were analyzed. Strategies to promote motivation, ensure safety, and define system requirements for remote communication and monitoring in VR telerehabilitation were also examined and will be discussed in the following sections.

The review included five studies categorized by methodology: pilot studies \cite{sramka2020, ferreira2020}, cohort studies \cite{Fregna2020}, and single-case studies \cite{salisbury2020, bailey2022}. The total sample size across these studies ranged from 1 to 56 participants, amounting to 95 individuals in total. Of these, 49 were stroke survivors with upper (47) or lower (2) limb impairments in subacute or chronic phases, while 46 were non-stroke participants who served as control groups. The studies were conducted in diverse locations, including Italy (1), the United States (2), Portugal (1), Slovakia and Czechoslovakia (1).

The studies aimed to evaluate several aspects of VR telerehabilitation for upper limb recovery in stroke survivors, including feasibility \cite{Fregna2020, salisbury2020, sramka2020, ferreira2020}, efficacy \cite{sramka2020, salisbury2020, bailey2022}, usability \cite{salisbury2020, ferreira2020, Fregna2020}, and adherence \cite{salisbury2020, bailey2022}. Notably, only three of the studies were conducted in home settings, while the remaining studies took place in clinical environments. Table \ref{tab:research-details} provides a detailed summary of the research findings.

\begin{table*}
  \caption{Summary of Research Details}
  \label{tab:research-details}
  \footnotesize
  \begin{tabular}{ p{1.5cm} p{1.2cm} p{1.2cm} p{3cm} p{8.7cm} }
    \toprule
    \textbf{Article} & \textbf{Country} & \textbf{Setting} & \textbf{Sample} & \textbf{Intervention Details} \\
    \midrule
   Fregna \textit{et al.} 2022 \cite{Fregna2020} &
   Italy &
   Clinical &
   16 subacute and chronic post-stroke patients (4 female, mean age 62 ± 9) &
   Study evaluated VR for upper limb motor rehabilitation in stroke patients. Patients had 50-minute sessions, showing high comfort and satisfaction. Significant motor function improvements correlated with clinical scores, suggest intervention effectiveness. \\
   \midrule
   Joseph P. Salisbury \textit{et al.} 2020 \cite{salisbury2020} &
   United States &
   Home &
   64-year-old male chronic patient &
   The CogniviveVR system, combining VR with therapeutic gaming, was used for post-stroke upper limb rehabilitation. Over eight weeks, the patient engaged in 30-minute sessions 5-6 times weekly. Outcomes showed improved movement smoothness and adherence, with 5,197 repetitions completed. \\
   \midrule
   Rachel B. Bailey, 2022 \cite{bailey2022} &
   United States &
   Home&
   75-year-old male chronic patient &
   Patient engaged in daily 30-minute VR sessions and traditional telehealth twice weekly for a month, then weekly for the second month. Over two months, he showed significant improvements in strength, and daily activities, with a Stroke Impact Scale score increase from 12 to 15 and a 90\% compliance rate\\
   \midrule
   Bruno Ferreira \textit{et al.} 2020 \cite{ferreira2020} &
   Portugal &
   Clinical &
   15-stroke patients, ages between 30 and 90 years &
   The intervention used an adaptive VR-based serious game for upper limb rehabilitation. Patients participated in 28-minute VR sessions, five times each. The Tetris-like game engaged and motivated patients, with measured outcomes showing adequacy in 75\% of clinical cases.\\
   \midrule
    Miron Sramka \textit{et al.} 2020 \cite{sramka2020} &
    Slovakia and Czechoslovakia &
    Home and clinical &
    6-stroke patients (2 only participated in lower limb scenarios) &
    The intervention combined VR sessions twice a week with traditional rehab once a week for six weeks. Focusing on limb and fine motor skills, patients showed significant motor skill progress and increased motivation.  \\
   \bottomrule
  \end{tabular}
\end{table*}

\subsection{How Do Patients and Clinicians Collaborate? (Q1)}

Building on the design principles and strategies discussed in Section \ref{subsec:HCD} and informed by \cite{brandon2019}, we analyzed the relevant literature on VR-based telerehabilitation. These guidelines provided a framework for structuring our analysis rather than evaluating study methodologies. This section examines recruitment and participant selection, journey mapping and goal setting, stakeholder roles, prototyping processes, and the integration of patient and expert feedback.

\subsubsection{Recruitment and Participant Selection}

The recruitment process for patients in the included studies considered various demographic and clinical factors, including age, sex, stroke type (ischemic or hemorrhagic), stroke phase (acute, subacute, or chronic), and the number of days elapsed since the event. In the study by \cite{Fregna2020}, severe cognitive impairments or other significant co-existing clinical conditions were ground for exclusion from the study. Different clinical evaluations of upper limb function were conducted across all studies, assessing balance \cite{sramka2020}, spatial perception, susceptibility to motion sickness \cite{sramka2020, Fregna2020}, motor skills \cite{sramka2020, bailey2022, Fregna2020, ferreira2020}, fine motor skills, and cervical spine mobility for 360-degree VR environments \cite{sramka2020}. Moreover, the presence of compensatory strategies and disease-related dysfunctions were assessed, particularly in chronic patients \cite{salisbury2020, bailey2022}. Issues reported included joint contractures in the affected hand and wrist \cite{bailey2022}, deficits in somatosensory feedback \cite{salisbury2020}, muscle weakness \cite{bailey2022, salisbury2020}, reduced active range of motion \cite{bailey2022}, and aphasia \cite{bailey2022}. Prior knowledge of VR was assessed in \cite{sramka2020, Fregna2020, ferreira2020}, as familiarity with the technology can facilitate navigation, and interaction with the system.

\subsubsection{Journey Mapping and Goal Setting}

While journey mapping was commonly addressed, details were often sparse. Initial patient assessments informed custom rehabilitation plans, as seen in Salisbury et al. \cite{salisbury2020} and Ferreira et al. \cite{ferreira2020}. However, patient involvement in early design stages was generally limited, with patients often unaware of tasks until testing \cite{salisbury2020}. During testing, tasks were adjusted to match patient performance \cite{ferreira2020, bailey2022, Fregna2020}. Only Fregna et al. \cite{Fregna2020} and Bailey et al. \cite{bailey2022} designed tasks closely aligned with patients' goals and daily activities, such as reaching for a glass or manipulating objects relevant to the patient's interests.

\subsubsection{Prototyping and Iterative Refinement}

Prototyping and iterative refinement, core to HCD, rely on ongoing feedback from users and experts. While the reviewed studies made system adjustments--primarily task modifications responding to patient progress or for future iterations \cite{Fregna2020, sramka2020}--explicit documentation of prototyping with ongoing user feedback was limited. Salisbury et al. \cite{salisbury2020} described developing prototypes based on specifications refined through expert interviews but lacked detailed accounts of the prototyping process and user feedback integration. This suggests that although iterative refinement was likely part of all studies, comprehensive documentation of these processes was generally absent, indicating a gap in reporting rather than in practice.

\subsubsection{Roles and Integration of Feedback}

Despite limited documentation of the prototyping process, integration of patient and expert feedback was evident across studies. Initial consultations were used to design custom environments \cite{sramka2020, Fregna2020} and tailor tasks to patient capabilities \cite{sramka2020, Fregna2020, ferreira2020, bailey2022, salisbury2020}, ensuring VR movements mirrored traditional rehabilitation. Feedback was collected during or after testing via self-report questionnaires such as the Stroke Impact Scale \cite{bailey2022} and the User Experience Questionnaire \cite{ferreira2020}, assessing various user experience factors. Additional feedback on experience, comfort, ownership, and agency was gathered \cite{Fregna2020}, and direct consultations via telehealth provided further insights \cite{sramka2020, salisbury2020}. Primary stakeholders included stroke survivors and clinicians (physiotherapists, occupational therapists). Psychologists were involved in participant selection in one study \cite{sramka2020}, while a neurologist referred a patient to VR therapy in another \cite{bailey2022}. Patients, as primary users, engaged with VR systems and provided experiential feedback. Therapists evaluated and selected patients, guided rehabilitation, adjusted plans based on progress, and monitored data through telehealth consultations \cite{bailey2022} or dedicated apps \cite{sramka2020, salisbury2020, Fregna2020, ferreira2020}.

\subsection{How is Monitoring and Communication Facilitated? (Q2)}

Monitoring and communication are critical components of any rehabilitation process, and their role becomes even more significant in the context of telerehabilitation. Unlike traditional rehabilitation, where therapists can directly observe and adjust treatments, telerehabilitation relies on remote monitoring to track patient progress, modify treatment plans, and mitigate risks of injury or treatment stagnation. This section examines how studies incorporated these features by exploring the types of data collected, tools and platforms used for remote monitoring and communication. Additionally, it explores how rehabilitation programs are adjusted based on monitoring, the challenges in transitioning to home-based rehabilitation, and examines how data security and patient privacy are managed.

\subsubsection{Data Collected for Patient Monitoring}

Remote monitoring employed technologies like VR headsets and motion tracking devices (e.g., Leap Motion) \cite{sramka2020} to collect quantitative data on hand positions, limb trajectories, and joint angles. Due to patient independence or system constraints, real-time data access was often impractical. Instead, data later was accessed via dashboards and apps \cite{salisbury2020, ferreira2020}. For instance, \cite{salisbury2020} recorded session logs capturing system usage, session length, repetitions, 3D motions, and movement smoothness. Moreover, game-based therapy elements were common across interventions, with metrics like response times, accuracy, task completion times, and gameplay parameters (e.g., scores, levels, playtime) collected to assess performance. In \cite{Fregna2020}, these metrics were correlated with clinical assessments like the FMA-UE, serving as indicators of therapeutic efficacy.

\subsubsection{Systems and Interfaces for Remote Rehabilitation Monitoring}

Various tools were used depending on the setting--home-based \cite{sramka2020, salisbury2020, bailey2022} or clinical \cite{Fregna2020, ferreira2020}. Home-based systems included VR headsets like HTC Vive, Oculus Rift, and Oculus Quest 2, or 2D monitors when HMDs were unsuitable \cite{ferreira2020}. Equipment choices were sometimes justified: Oculus Quest 2 for its specifications and affordability \cite{Fregna2020}, Oculus Rift for validated kinematic measurements \cite{salisbury2020}. Therapist interaction varied; in \cite{bailey2022}, VR sessions were interspersed with traditional therapy and monitored via telehealth. In \cite{sramka2020}, remote exercises were managed through online sessions with additional activity tracking. In \cite{salisbury2020}, patients self-administered treatment with data shared via a clinician dashboard. As for the systems tested in clinical settings, they featured additional components, though their efficacy was not thoroughly assessed due to the close supervision by clinical staff. For instance, \cite{Fregna2020} allowed therapists to interact vocally with patients via a wireless connected PC providing a detailed VR render, and \cite{ferreira2020} developed a system comprising a gaming application, central database, and therapist back-end application, with the latter allowing therapists to review data from previous sessions, adjust game difficulty, and update patient information

\subsubsection{Adapting Rehabilitation Based on Monitoring Results}

All systems incorporated mechanisms to adapt treatment based on monitoring results. In studies with therapist presence during sessions \cite{sramka2020, bailey2022}, patient progress was closely monitored, and tasks were adjusted with real-time feedback. In studies without direct supervision \cite{salisbury2020, Fregna2020, ferreira2020}, adaptation was achieved through performance metrics enabling therapists to modify session parameters, such as task sequences and difficulty levels. Some systems employed Dynamic Difficulty Adjustment (DDA) to automatically adjust activity difficulty based on patient performance \cite{salisbury2020, ferreira2020}.

\subsubsection{Barriers and Challenges in Transitioning to Home-Based Rehabilitation}

The studies identified several barriers and difficulties that could affect the successful transition to telerehabilitation. One major concern was technological accessibility. The shift to home rehabilitation requires patients to have a basic proficiency with VR devices \cite{sramka2020}. Initial sessions often focus on training patients and caregivers on how to use the VR headset, which can be challenging for those who are not tech-savvy or living alone \cite{bailey2022}. Additionally, \cite{salisbury2020} highlighted that external hardware and the need for a separate computer might complicate and increase the cost of at-home deployment. Therapists noted limitations in remote assessment, such as inability to measure strength and muscle tone remotely \cite{bailey2022}, which are routinely performed in-person. Lastly, data privacy was also a concern for broader implementations outside secure research environments \cite{Fregna2020}.

\subsubsection{Managing Data Security and Patient Privacy}

Only \cite{Fregna2020} addressed data security and patient privacy in detail, proposing a plan for future deployments. Measures included not storing personal information on VR headsets (patients identified by codes), securely maintaining the link between codes and personal information on therapists' PCs protected by security protocols, and securing the connection between client apps and HMDs with unique keys to authorize access and ensure controlled offline access.

\subsection{How are Well-being and Safety Assured? (Q3)}

Ensuring patient safety is paramount in VR-based telerehabilitation due to the unique challenges posed by virtual environments and remote interactions. This section evaluates the measures implemented to safeguard patients, focusing on risk mitigation strategies, management of potential adverse effects, and protocols for emergency situations and technical failures.

\subsubsection{Risk Mitigation Strategies and Safety Protocols in VR-Based Telerehabilitation}

Safety measures varied across studies, reflecting different identified risks such as physical strain, visual strain, and motion sickness. A common approach was the continuous clinician support before and during treatment, with regular evaluations to adjust rehabilitation scenarios to patient needs. In \cite{sramka2020, bailey2022, Fregna2020}, therapists were available online during sessions, offering contact and vigilance similar to traditional therapy. This setup allowed for real-time monitoring of patient movements and feedback to ensure exercises were performed correctly and safely. However, this approach limits one of the potential advantages of telerehabilitation, which is the ability to simultaneously reach and manage a broader patient base. To address this, \cite{ferreira2020} developed a system supporting multiple play-stations, allowing therapists to monitor multiple patients simultaneously. Conversely, \cite{salisbury2020} relied on post-session calibration and performance data to adjust treatment, utilizing a therapeutic framework combining DDA with clinically validated approaches to ensure tasks were challenging yet within physical limits.

Preemptive safety measures included training patients and caregivers on VR headset use \cite{salisbury2020, bailey2022, ferreira2020}, ensuring proper home setup \cite{bailey2022}, and verifying communication between devices. To mitigate motion sickness, also known as cybersickness, some systems were calibrated to be used in beds \cite{salisbury2020} or chairs \cite{Fregna2020}, reducing fall risk and minimizing visual-vestibular conflicts. Fregna et al. \cite{Fregna2020} further addressed motion sickness by limiting session durations and enhancing sensorimotor congruence through tactile and auditory feedback, such as aligning real objects with their virtual counterparts. Additionally, \cite{ferreira2020} carefully positioned objects in the virtual environment to promote visual exploration while avoiding visual strain.

\subsubsection{Adverse Effects and Protocols for Technical Failures and Emergency Situations}

No adverse events were reported in the reviewed studies. In \cite{bailey2022}, a technical issue arose due to sensor misalignment after changing the placement of Oculus sensors. This was resolved remotely by guiding the participant's family over the phone, avoiding the need for an in-home visit. Regarding protocols for technical failures, only \cite{ferreira2020} had a structured approach for loss of connectivity. Their system supports a temporary offline mode where data is saved locally and later uploaded to the database when connectivity is restored. However, backend support, including progress reports and game modifications by therapists, still requires an active internet connection.

\subsection{How is Engagement Promoted? (Q4)}

Stroke survivors often face decreased motivation and adherence to recovery programs after hospitalization, slowing their recovery. VR offers a promising solution by enhancing engagement and motivation in rehabilitation. This section examines how reviewed studies utilized VR to promote patient engagement, focusing on strategies to boost adherence and the design of exercises within VR systems.

\subsubsection{Strategies for Fostering Motivation and Engagement}

In exploring strategies to promote patient motivation and engagement in VR-based rehabilitation, the reviewed literature highlights personalization as a key approach to addressing individual needs and preferences. One notable method is leveraging the "WOW" effect to sustain interest and motivation \cite{sramka2020, Fregna2020, salisbury2020}. For instance, \cite{sramka2020} allowed patients to select virtual environments enriched with multisensory inputs, such as fragrance, music, and lighting, to evoke positive emotions. Similarly, \cite{Fregna2020} created a "cozy" VR environment by calibrating virtual elements to match real-world objects, fostering embodiment, presence, proprioception, and immersion through sensorimotor congruence.

Gamification was another common strategy, transforming therapeutic exercises into playful activities to encourage prolonged engagement. Game-based therapies introduced short-term goals and rewards to foster a sense of progress and accomplishment. For instance, \cite{salisbury2020} developed a minigame where patients hit a ball with a virtual paddle, incorporating levels, rewards (i.e., 1-3 stars), and DDA to tailor challenges to patient performance. Similarly, \cite{bailey2022} created applications like "\textit{Balloon Blast}" and "\textit{Color Match}", integrating physical and cognitive challenges. In \cite{Fregna2020}, the design featured tasks like "\textit{Ball in Hole}", "\textit{Cloud}", "\textit{Glasses}", and "\textit{Rolling Pin}" engaging different hand movements and coordination. Finally, \cite{ferreira2020} introduced a Tetris-like game with adaptive difficulty to balance challenge and engagement, easing the learning curve for inexperienced users.

\subsubsection{Targeted Motor Skills}

The VR-based rehabilitation systems included exercises targeting various motor skills. To promote fine motor skills, tasks required precise hand movements, such as manipulating virtual objects with hand gestures \cite{sramka2020}, and tasks like "\textit{Ball in Hole}" and "\textit{Glasses}" \cite{Fregna2020}. The Tetris-like game by \cite{ferreira2020} encourage hand coordination to manipulate falling shapes. For compound movements involving multiple limbs, exercises like "\textit{Balloon Blast}" \cite{bailey2022} improved shoulder range of motion through swiping motions, while "\textit{Rolling Pin}" \cite{Fregna2020} required both hands to promote gross motor skills. Some exercises also stimulated visual and cognitive skills; for example, \cite{ferreira2020} included tasks requiring visual exploration, promoting visuospatial training through larger head and gaze rotations.

\section{Discussion}

This systematic review examined the current landscape of VR-based telerehabilitation systems, categorizing their characteristics by addressing key questions on design principles, facilitation of monitoring and communication, mitigation of safety concerns, and strategies to promote patient motivation and engagement. Our aim was to provide a comprehensive overview of factors contributing to the success of VR-based telehealth interventions.

\subsection{From Participation to Play}

A significant focus was placed on the adoption of HCD principles in developing VR games and systems for telerehabilitation. We aimed to delineate stakeholder roles, describe their involvement in design and development processes, and explore how their contributions shaped outcomes, guided by the design strategies proposed by \cite{brandon2019}.

Our review identified a notable gap in the literature regarding methodologies for system development, particularly in stakeholder definition and involvement during design phases, as noted by previous studies \cite{Rodrigues2024}. Most studies concentrated on later stages like user testing and validation, with limited early-stage involvement from stakeholders—especially patients. For instance, \cite{sramka2020} involved clinicians in initial consultations, but patient involvement in early design stages was minimal. Therapists generally played more significant roles in guiding rehabilitation, adjusting plans based on progress, and monitoring patient data, but their contributions were primarily during later stages. While customization and adaptation to patient capabilities and therapeutic goals were emphasized, detailed reporting of these processes was lacking, as evidenced by the absence of explicit journey mapping.

There was also a substantial gap in reporting the prototyping process and integration of feedback from key stakeholders. Studies like \cite{salisbury2020} mentioned prototype development through expert consultation but lacked detailed descriptions of the actual prototyping process and how user feedback was incorporated. Although most studies reported some collaborative efforts and refinements based on stakeholder input \cite{sramka2020, Fregna2020, ferreira2020, salisbury2020}, these primarily focused on adapting systems according to patient progress and proposing changes for future iterations, without detailing the HCD process. This omission makes it challenging to correlate stakeholder involvement with outcomes and understand how iterative refinement was conducted.

HCD fundamentally operates as a bottom-up approach, where aspects like safety, technology, satisfaction, well-being, and efficacy are established through its principles. By prioritizing end-user needs and inputs, HCD ensures systems are developed to address the most critical elements of user experience and clinical efficacy. This highlights HCD's holistic nature in designing every component with the user's needs at the forefront. Therefore, detailed reporting from initial stakeholder involvement to iterative prototyping is critical. The lack of detailed descriptions of decision-making in early development phases creates a disconnect between system design and outcomes, compromising reproducibility. This makes it challenging for other researchers and developers to replicate studies or interventions accurately. Comprehensive documentation is essential for transparency in the design and development process, showing how decisions were made, how feedback was considered, and how the system evolved. Such transparency fosters scientific rigor, accountability, and contributes to a robust body of knowledge and best practices in HCD and VR interventions.

\subsection{The Art of Monitoring and Communication}

Monitoring and communication are essential in telerehabilitation for continuous tracking of patient progress and adjustment of rehabilitation plans. This review highlights various technologies used in VR-based telerehabilitation, offering different levels of patient independence and therapist involvement. Remote monitoring in these systems primarily relies on commercial VR headsets and motion tracking devices, which provide detailed quantitative data on patient movements such as hand positions, limb trajectories, response times, and accuracy. This data is used for assessing therapeutic efficacy and tracking patient progress.

There is a dichotomy in patient independence among systems. Solutions like those in \cite{sramka2020} and \cite{bailey2022} involve direct therapist supervision through telehealth sessions, allowing real-time adjustments and immediate feedback but potentially limiting patient autonomy. Conversely, systems like those in \cite{salisbury2020} and \cite{Fregna2020} enable patients to self-administer treatments with therapists reviewing data periodically, promoting independence but possibly reducing the immediacy of therapeutic adjustments.

While both approaches have merits, they share significant gaps. Specifically, neither fully leverages VR technology's capabilities to provide comprehensive sensory integration and haptic feedback, which are crucial for effective stroke rehabilitation. VR offers significant advantages by providing sensory inputs that facilitate motor relearning and promote neuroplasticity; however, achieving comprehensive sensory integration often requires instrumented or passive objects to provide haptic feedback. Only one study, by \cite{Fregna2020}, attempted to enhance the sense of embodiment and proprioception by calibrating the virtual environment to match a physical object. While this approach improves immersion, it highlights another limitation and common concern among therapists: the lack of physical contact, which is standard for assessing patient well-being and comprehensively understanding patient progress.

A potential strategy to bridge the gap between telerehabilitation and conventional therapies is to integrate existing physiotherapy devices into the VR rehabilitation process. Traditional physiotherapy employs auxiliary devices such as supination/pronation grippers, rope and pulleys, and shoulder wheels to support task execution and recover joint movement amplitude \cite{ferreira2020}. Adapting these devices for use in a VR environment could enhance their functionality and provide additional performance data. For example, shoulder wheels could be fitted with sensors to track rotations or measure partial spins, while instrumented rope pulleys and grippers could provide data on movement speed, amplitude, and strength. Incorporating additional sensor data, such as physiological metrics like electromyography (EMG), could offer deeper insights into muscle activity, fatigue, and compensatory mechanisms \cite{patricia2022}. This integration would address the physical contact limitations of telerehabilitation and improve the therapeutic process by providing therapists with richer, quantitative measures of efficacy.

Finally, ensuring robust data security and patient privacy is critical. While the reviewed studies were conducted in research settings, future home or non-clinical deployments must consider regulatory compliance and the protection of patient information.

\subsection{Safety First}

The exploration of safety concerns in telerehabilitation and VR rehabilitation uncovers a complex interplay between the benefits of remote care and the inherent risks associated with it. As telerehabilitation shifts from traditional in-person supervision to remote modalities, patient safety and well-being remain paramount, introducing new challenges that must be carefully managed to prevent injury.

To promote safety, various strategies have been employed across different systems. For instance, continuous clinician support, both preemptively and during treatment, allows for regular adjustments tailored to patient needs. Systems that facilitate real-time therapist presence during sessions---such as those described by \cite{sramka2020}, \cite{bailey2022}, and \cite{Fregna2020}---enable immediate monitoring and feedback. This helps ensure proper exercise execution and mitigates the risk of injury by allowing therapists to make on-the-fly corrections and provide guidance.

Addressing issues of workload management, systems like those of \cite{salisbury2020, ferreira2020}, employ custom DDA, which adjusts task difficulty in real-time to prevent frustration and undue physical strain, thereby reducing the risk of overexertion and associated injuries. Additionally, preemptive safety measures such as training for patients and caregivers on safe VR headset use and ensuring proper home setup are critical in mitigating risks related to unsupervised environments and improper equipment use \cite{salisbury2020, bailey2022, ferreira2020}.

Motion sickness, a common concern in VR environments, is managed through various approaches. Systems are calibrated for use in stable settings like beds or chairs, minimizing the risk of falls and reducing visual-vestibular conflicts \cite{salisbury2020, Fregna2020}. Limiting session durations and enhancing sensorimotor congruence with tactile and auditory feedback further helps to reduce visual and physical strain, and contributes to a safer VR experience \cite{Fregna2020}.

Despite these measures, certain concerns remain inadequately addressed. A significant gap is the lack of quantitative channels to assess a patient’s physical capabilities, which increases the risk of musculoskeletal injuries, particularly since immediate intervention during remote sessions is not possible if patients experience discomfort or adverse reactions. The reviewed studies did not provide emergency protocols for such scenarios. For instance, in \cite{salisbury2020}, a technical problem required the participant’s family to be guided over the phone, which wouldn't work for those living alone and without constant caregiver support.

Preemptive strategies are necessary in mitigating these risks. It would be beneficial for future research to provide more detailed and comprehensive approaches of addressing safety concerns, particularly for home or non-clinical deployments.

\subsection{The Joy of Play}

Integrating game-based therapies into post-stroke rehabilitation addresses the monotonous and physically demanding nature of traditional protocols by introducing a more engaging approach. This method provides valuable quantitative feedback while transforming rehabilitation into a playful experience that enhances patient adherence and motivation. VR is particularly advantageous in this context, offering a broad range of sensory stimuli and precise tracking of patient movements, which facilitates continuous monitoring and increases engagement through immersive experiences.

Our review highlights the considerable potential of these strategies, especially in settings with limited clinical supervision such as telerehabilitation, where maintaining patient motivation can be challenging. Personalization and gamification emerge as key strategies, incorporating reward systems and short-term goals into therapeutic exercises to encourage consistent performance. The use of DDA further enhances engagement by adapting task difficulty in real time based on patient performance, ensuring challenges are appropriately matched to skill levels. These gamified approaches not only make rehabilitation more enjoyable but also provide a more interactive and personalized experience, helping patients stay aware of their progress and remain motivated throughout their recovery.

Despite progress in VR-based rehabilitation, there remains a lack of consensus regarding the terminology used to describe various VR interventions. Terms such as "activities," "minigames," "games," and "serious games" are often used interchangeably without clear differentiation. Furthermore, many current VR applications exhibit simplistic characteristics, focusing primarily on first-person virtual experiences that replicate real-life activities or engage users in coordinated movements—implying that participants possess a baseline level of strength. This limited approach fails to fully exploit advanced game design principles and narrative elements that could enhance patient agency and autonomy, potentially leading to more effective therapeutic experiences.

Exploring designs that promote social engagement with family members or other patients, and incorporating personal goals aligned with previously enjoyed recreational activities, could provide additional motivation and support \cite{lorenz2024}. For example, interactive designs allowing patients to engage with family or connect with others online could foster a sense of community. Integrating patients' personal goals into the VR experience, broken down into manageable steps, could further challenge and motivate them, facilitating functional improvement \cite{bailey2022}.

Finally, the long-term sustainability of these VR systems was not a primary focus in the reviewed studies. Many systems rely on the initial "WOW" effect to engage users, particularly older adults new to VR technology. However, as users become more familiar with the system, this initial excitement may diminish, potentially leading to decreased engagement over time.

\section{Conclusion}

This systematic review mapped the current landscape of VR-based telerehabilitation systems by examining their design principles, monitoring and communication methods, safety concerns, and strategies to increase patient motivation and engagement. Our findings highlight the potential of VR interventions to offer engaging, adaptive, and personalized telerehabilitation experiences. However, several challenges require further attention.

A key finding is the limited application of HCD principles during early development stages. While many studies emphasized user testing and validation, they lacked documentation of early design decisions and stakeholder involvement. This lack of transparency hampers reproducibility and may lead to a disconnect between the systems and the patients they are intended to serve.

Monitoring and communication systems varied in patient autonomy and therapist involvement. Although each system has its advantages, there is room for innovation—particularly in developing tools that improve the therapeutic experience and performance evaluations in home settings. Challenges such as the lack of standardized frameworks, limited safety protocols for remote care—especially concerning emergency response and quantitative assessments of physical capabilities—and data privacy concerns need urgent attention. Developing robust guidelines and addressing these gaps will be critical for broader clinical adoption.

Game-based therapies in VR telerehabilitation have the potential to enhance patient adherence and motivation through personalization and gamification. However, current applications, while ecologically valid, often do not incorporate advanced game design, narrative elements, or social interactions that could further promote patient agency and autonomy.

The significant variability among studies, with differing priorities and solutions, makes it challenging to identify consistent success factors for VR telerehabilitation. This underscores the need for better documentation of design and implementation processes to effectively correlate system effectiveness with specific design principles. Comprehensive documentation would enable more reliable comparisons of interventions, evaluating them not just for effectiveness but also for how they address key challenges. Without standardization, comparisons between interventions may be inequitable, as those addressing a broader range of issues offer a more holistic perspective.

With ongoing innovation, VR telerehabilitation holds the potential to revolutionize stroke recovery and redefine patient care. However, significant challenges remain, and further advancements are needed before achieving broader clinical adoption and long-term impact.

\section{Limitations}

This review has several limitations that should be acknowledged. First, the limited number of studies included restricts the generalizability of our findings. The scarcity of research focusing specifically on VR-based telerehabilitation for upper limb recovery post-stroke highlights the nascent stage of this field and underscores the need for more comprehensive studies. Second, there was significant variability among the included studies in terms of design methodologies, intervention protocols, and outcome measures. This heterogeneity made it challenging to draw consistent conclusions or identify definitive success factors for VR telerehabilitation interventions. The lack of standardized methodologies and reporting practices further complicates the comparison of results across studies and hinders the synthesis of evidence. Finally, the rapid evolution of VR technology means that some of the interventions reviewed may already be outdated, limiting the applicability of the findings to current and future developments in the field.

\bibliographystyle{vancouver}

\end{document}